\documentclass[seceq,supplement]{ptptex}

\usepackage{graphicx}
\usepackage{wrapft}
\graphicspath{{fig/}}

\title{
Equation of State of Structured Matter at Finite Temperature%
}

\author{
Toshiki \textsc{Maruyama}$^1$,
Nobutoshi \textsc{Yasutake}$^2$
and 
Toshitaka \textsc{Tatsumi}$^3$%
}

\inst{
$^1$Advanced Science Research Center, Japan Atomic Energy Agency,\\
Shirakata Shirane 2-4, Tokai, Ibaraki, 319-1195 Japan\\

$^2$Division of Theoretical Astronomy, National Astronomical Observatory of Japan, 
2-21-1 Osawa, Mitaka, Tokyo, 181-8588 Japan\\

$^3$Department of Physics, Kyoto University, Kyoto, 606-8502 Japan
}

\abst{

We investigate the properties of 
nuclear matter at the first-order phase transitions such as 
liquid-gas phase transition and hadron-quark phase transition.
As a general feature of the first-order phase transitions of matter consisting of many species of
charged particles,
there appears a mixed phases with 
geometrical structures called ``pasta'' due to the balance of the Coulomb repulsion
and the surface tension between two phases.\cite{Ravenhall,hashimoto}
The equation of state (EOS) of mixed phase is different from the one obtained 
by a bulk application of the Gibbs conditions or by the Maxwell construction
due to the effects of the non-uniform structure.
We show that the charge screening and strong surface tension make
the EOS close to that of the Maxwell construction.
The thermal effects are elucidated as well as the above finite-size effects.
}

\begin{document}

\maketitle

\section{Introduction}

There are several phase transitions in nuclear matter, some of which
are of the first order.
It is well-known that there appears the mixed phase during the first-order phase transitions.
If the system consists of single chemical component, the Maxwell construction
satisfies the Gibbs conditions, i.e.\ balance of pressure and chemical potentials 
among two coexisting phases.
In the case of nuclear matter, however, which consists of several independent
chemical components, one cannot apply the Maxwell construction.
Furthermore, the mixed phase shows a series of geometrical structures due to 
the Coulomb interaction between charged components in the system.
To minimize the sum of the Coulomb and the surface energies, the shape of 
the mixed phase changes from spherical droplet, rod, slab, tube, and to bubble
depending on the volume fraction of the coexisting phases.
According to the geometrical shapes, the structures of matter are 
often called as ``pasta'' structures.

We have developed a framework to study the pasta structure and calculate the equation of state (EOS), 
taking into account the effects of the Coulomb repulsion and the surface tension in a self-consistent way. 
In this scheme the charge screening effects is automatically considered. 
In the series of study we have investigated 
the liquid-gas mixed phase at sub-nuclear densities \cite{marupasta},
the kaon condensed mixed phase \cite{marukaon},
and the hadron-quark (HQ) mixed phase \cite{endo,yasutake}.
Then we have found that 
the Coulomb screening by the rearrangement of charged particles
reduces the Coulomb energy of the system and consequently
enlarges the size of the structure.
By the screening the local charge density decreases
so that the EOS of the mixed phase approaches 
that of the Maxwell construction.
Especially for the kaon condensation and the HQ transition 
the effects of charge screening are pronounced.
We have found a novel phenomenon, hyperon suppression 
in the HQ mixed phase \cite{maru07,maru08}.
All of the above results are at zero temperature.
For the stellar objects, however,
zero-temperature corresponds only to cold neutron stars.
In contrast, the collapsing stage of supernovae, 
proto neutron stars and neutron star mergers, 
which represent more vivid scene, 
are so warm as several tens MeV.

In this paper we investigate the structure and the EOS of
the mixed phase at finite temperature.
Particularly we are interested in whether there is
any difference between finite and zero-temperature cases.
In the following we concentrate our target to the low-density 
nuclear matter where the liquid-gas (LG) mixed phase is relevant.
Then we extend our discussion to the HQ mixed phase
which may be crucial for the properties of proto neutron stars.

\section{Liquid-gas mixed phase of nuclear matter}

First we investigate the property of low-density nuclear matter, charge neutralized by electrons.
We employ the relativistic mean field (RMF) model to describe the
properties of nuclear matter under consideration.
The RMF model with fields of 
$\sigma$, $\omega$ and $\rho$ mesons and baryons introduced
in a Lorentz-invariant way is not only relatively simple for 
numerical calculations, but also sufficiently
realistic to reproduce bulk properties of finite nuclei
as well as the saturation properties of nuclear matter \cite{marupasta,marurev}.
One characteristics of our framework is that
the Coulomb interaction is properly included in the 
equations of motion for nucleons, electrons and meson mean fields.
Thus the baryon and electron density profiles, as well as the meson
mean fields, are determined in a fully
consistent way with the Coulomb interaction.

To solve the equations of motion for the fields numerically,
we divided the whole space into equivalent Wigner-Seitz cells with 
geometrical symmetry. 
The shapes of the cell are 
sphere in three-dimensional (3D) cases, cylinder in 2D and slab in 1D.
Each cell is globally charge-neutral and all physical quantities
in the cell are smoothly connected to those of the next cell
with zero gradients at the boundary.
The coupled equations for fields in a cell are solved by a relaxation method
for a given baryon-number density 
under constraints of the global charge neutrality.
Parameters included in the RMF model are chosen to reproduce the saturation properties
of symmetric nuclear matter, 
and the binding energies and the proton fractions of nuclei.
Details of the parameters are explained in Refs.\ \citen{marupasta,marurev}.

When we study nuclear matter at finite temperature,
the momentum distribution function is
a Fermi-Dirac distribution instead of a step function with a threshold 
at the Fermi energy.
In the numerical calculation,
density, scalar density, and kinetic energy density, etc of a fermion $a$ 
are obtained by integrating the functions of $T$, $\mu_a$ and $m_a^*$
over all the momentum-space.
We store those values in tables and get necessary quantities by
interpolating them.
The finite-temperature effects on the meson fields and
the contribution of anti-particles are neglected for simplicity.

\begin{figure}
\centerline{
\includegraphics[width=.28\textwidth]{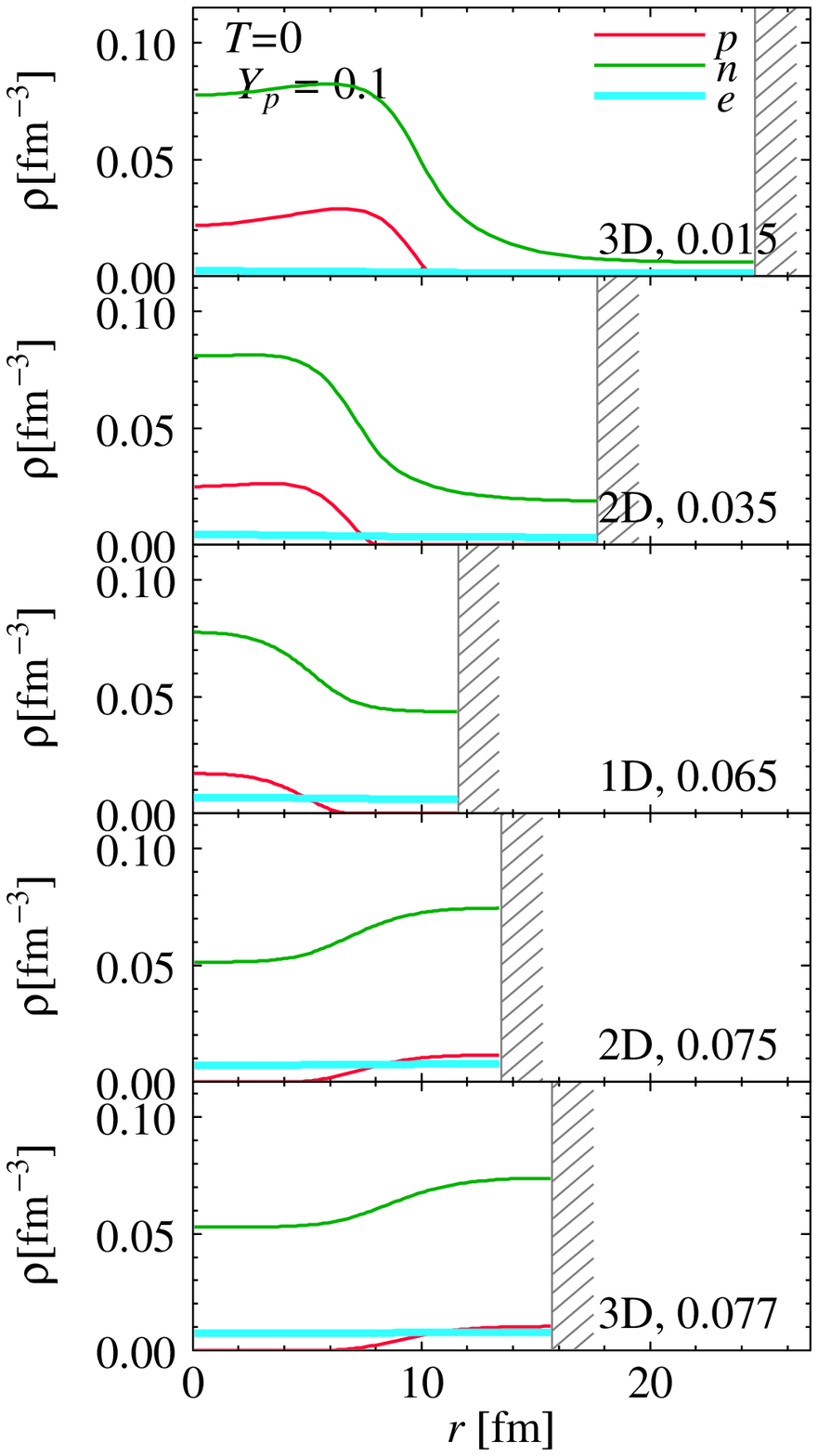}
\includegraphics[width=.28\textwidth]{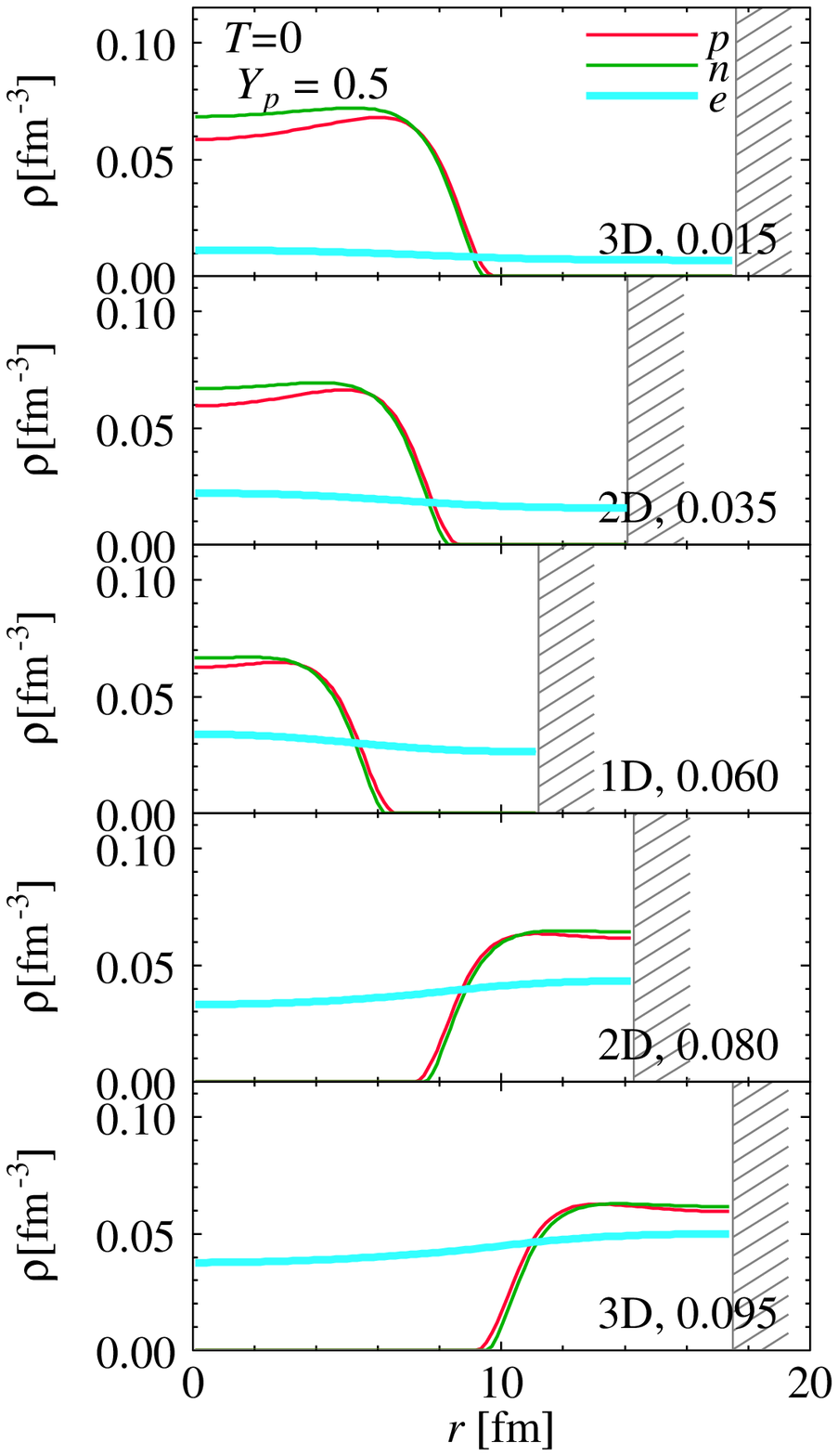}
\includegraphics[width=.28\textwidth]{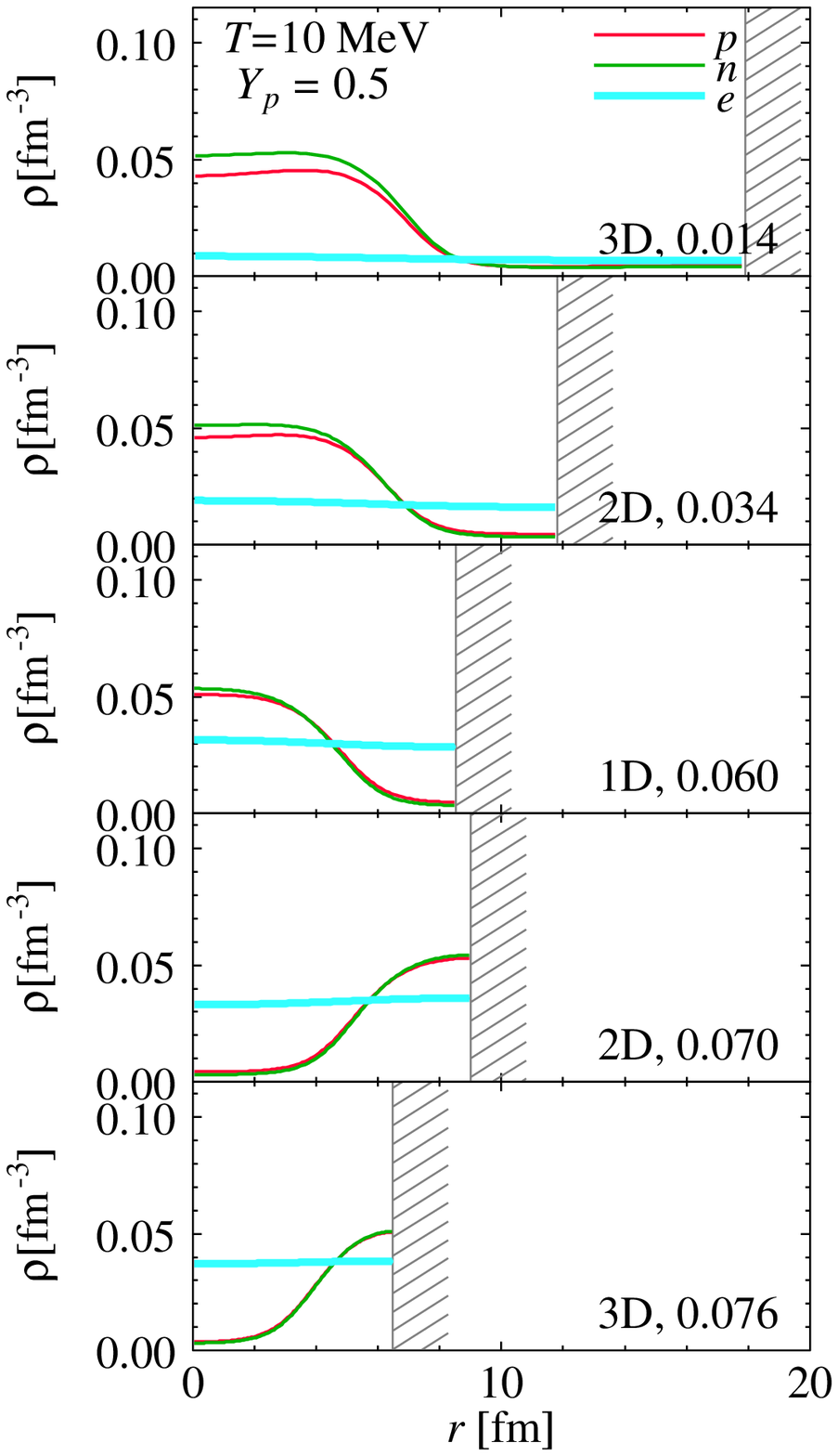}
}
\caption{
Examples of the density profiles in the Wigner-Seitz cells.
with $Y_p=0.1$ (left), $Y_p=0.5$ $T=0$ (center) and  $T=10$ MeV (right).
\vspace{-5mm}
}
\label{figPasta}
\end{figure}

First, we show in Fig.~\ref{figPasta} some typical density
profiles in the cell. 
The left and the central panels show the cases of proton fraction $Y_p=0.1$ and $Y_p=0.5$.
Apparently, dense nuclear phase (liquid) and dilute nuclear/electron phase (gas) 
are separated in space and they form pasta structures depending on density.
One should notice that coexisting two phases have
different components, i.e.\ nuclear matter and electron gas.
Therefore the EOS of the whole system cannot be obtained
by the Maxwell construction.
Since electron density is almost uniform 
and independent of baryon distribution,
we can discuss properties of the baryon partial system.

\begin{figure}
\centerline{
\includegraphics[width=.42\textwidth]{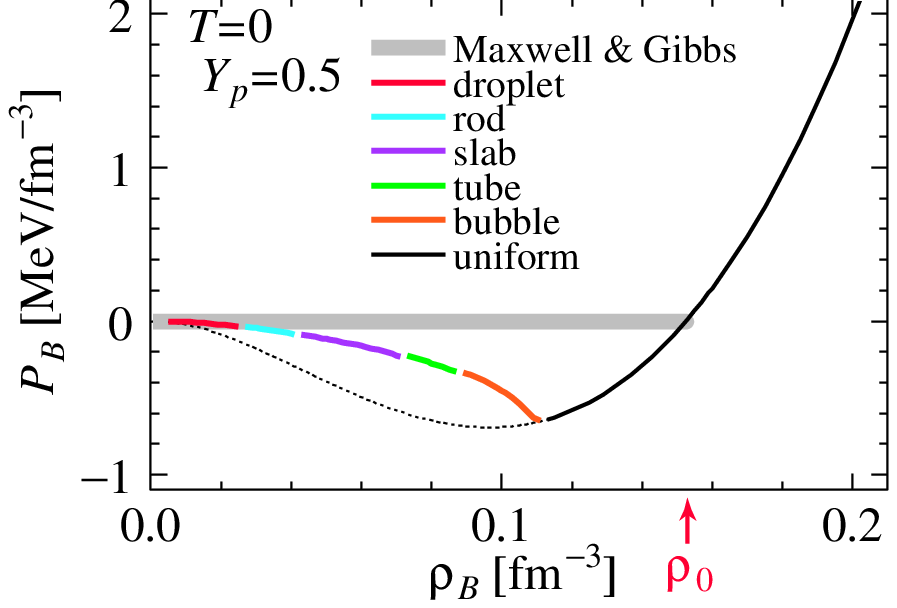}
\includegraphics[width=.42\textwidth]{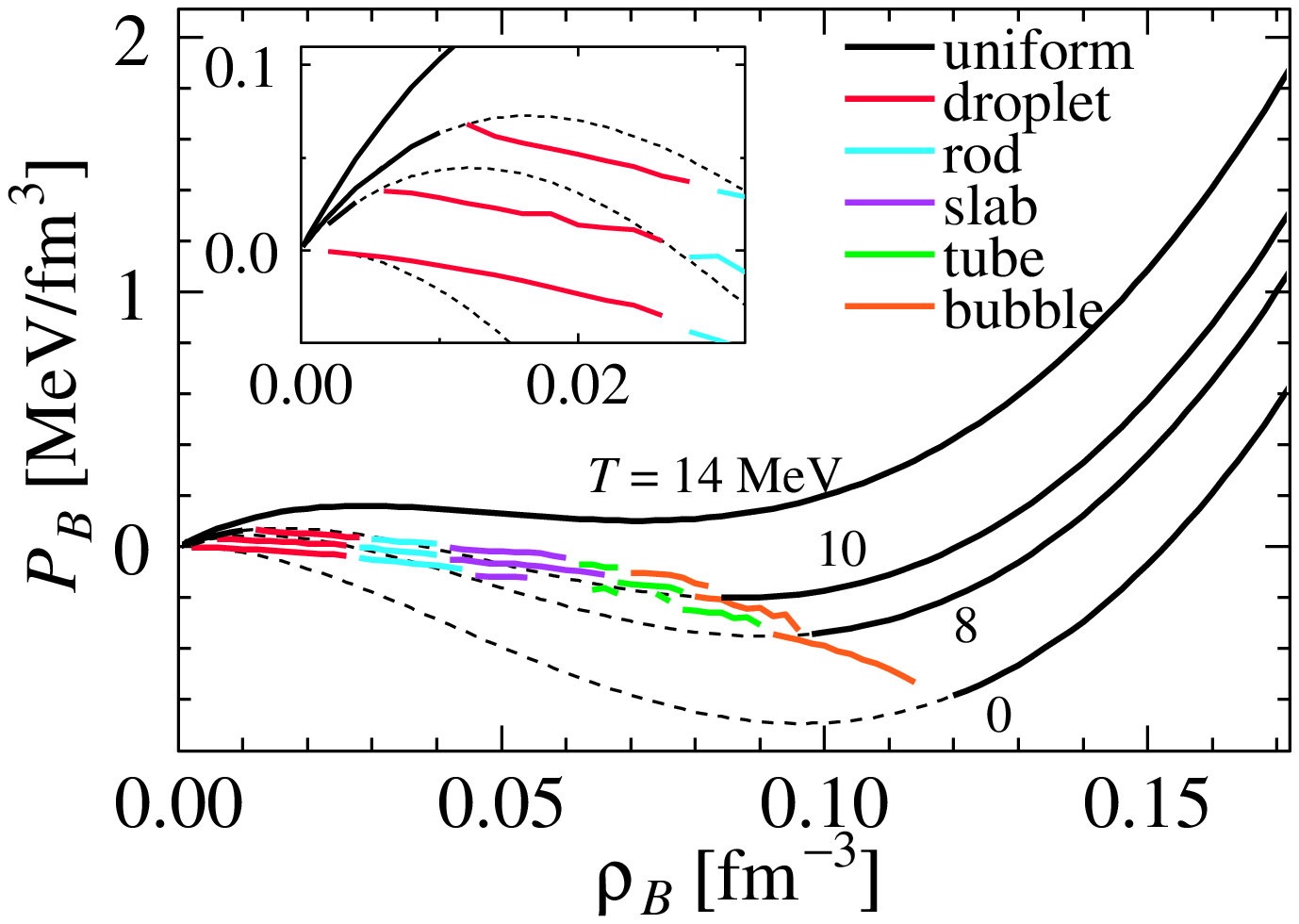}
}
\caption{
The baryon partial pressure as a function of density $\rho_B$
for symmetric nuclear matter $Y_p=0.5$.
\vspace{-5mm}}
\label{figEOS05}
\end{figure}

In the case of $Y_p=0.5$, 
the local proton fraction stays almost constant.
Therefore the system behaves like a system with single component.
This means that one can apply the Maxwell construction to get 
the baryon partial pressure $P_B$ as in the left panel of Fig.~\ref{figEOS05}:
uniform low-density matter with a negative partial pressure is 
not favored and the Maxwell construction gives $P_B=0$ for the mixed phase. 
By finite-size effects, i.e.\ the Coulomb repulsion and the surface tension,
structured mixed phase becomes unstable in the density region just below $\rho_0$,
and consequently uniform matter with a negative partial pressure is allowed.

In the case of asymmetric nuclear matter, 
e.g.\ $Y_p=0.1$ in the left panel of Fig.~\ref{figPasta},
the proton fraction in the dilute and dense phases are 
different, especially for low $Y_p$.
Matter behaves as a system with multi chemical components
and the Maxwell construction does not satisfy the Gibbs conditions.

Next, let us discuss about the thermal effects on the 
LG mixed phase of low density nuclear matter.
By comparing the density profiles in the central and right panels
of Fig.~\ref{figPasta},
we easily notice that 
the dilute phase at finite temperature always contains baryons 
while dilute phase at zero temperature is baryonless if $Y_p\approx 0.5$.
This is due to the Fermi distribution at finite temperature, where 
density as a function of chemical potential is always positive.

We also notice that the size of the pasta structure is smaller
in the case of finite temperature.
This comes from a reduction of the surface tension between two phases
at finite temperature since the difference of baryon density 
between two phases is smaller.

The EOS (baryon partial pressure as a function of baryon number density)
of symmetric nuclear matter 
at various temperatures is shown in the left panel of Fig.~\ref{figEOS05}.
Dotted and thick solid curves show the cases of uniform matter,
while thin solid curves are the cases where non-uniform structures
are present.
As shown in the right panel of Fig.~\ref{figEOS05},
pasta structures appear at finite temperatures as well as the case of $T=0$.
But there appears uniform matter (gas phase) at the lowest-density region
\cite{avancini,friedman}
since the baryon partial pressure of uniform matter has
a positive gradient against density.
On the other hand, the uniform matter is unstable
where the pressure gradient is negative
even if the pressure itself is positive.
At $T=14$ MeV, we obtain no pasta structure
since the baryon partial pressure of uniform matter
is a monotonic function of density.

\section{Hadron-quark mixed phase}

In this section 
we study the hadron-quark (HQ) phase transition with the finite-size effects at finite temperature.
For the hadron phase, we adopt  a realistic equation of state in the framework of the Brueckner-Hartree-Fock (BHF)
theory including hyperons.
For the quark phase, we use the MIT bag model with the bag constant $B$ for simplicity.
The MIT bag model is a successful one to reproduce the hadron mass spectrum, 
but may be an oversimplified model for quark matter.  
It should be interesting to compare our results with those given by other models; 
e.g.\ NJL model~\cite{burgio08,yasutake09} or PNJL model~\cite{fukushima08}.  
A sharp boundary is assumed between two phases in the HQ mixed phase and the surface energy
is taken into account in terms of a surface-tension parameter $\sigma_{\rm surf}$.
The details of calculation and interaction parameters are 
written in Ref.~\citen{maru07}.
The extension to finite temperature can be done with the same procedure   
as in the previous section.
The interaction energy of baryons and quarks are assumed to be independent on temperature.

\begin{figure}
\centerline{
\includegraphics[width=.41\textwidth]{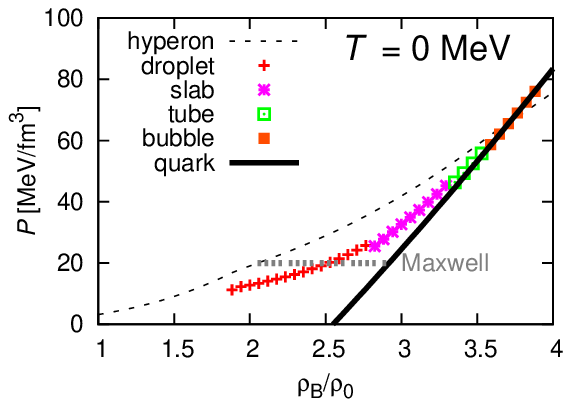}
\includegraphics[width=.41\textwidth]{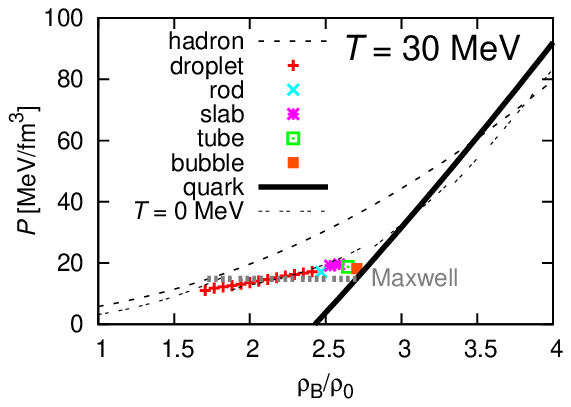}
\vspace{-4mm}
}
\caption{
The pressure of the mixed phase~(thick dots) in comparison with pure hadron and quark phases~(thin curves).
The left panel shows the case of zero temperature, and the right $T=30$~MeV.
We also show, for comparison, the mixed phase 
by the Maxwell construction by thin solid line.
\vspace{-5mm}
}
\label{fig:eos}
\end{figure}

Figure~\ref{fig:eos} shows
the resulting pressures of the HQ mixed phase with that of the pure hadron and quark phases
over the relevant range of baryon density at zero temperature (left panel) 
and finite temperature $T=30$ MeV (right panel). 
The thin curves indicate the pure hadron and quark phases,
while the thick dots indicate the mixed phase 
in its various geometric realizations by the full calculation.
We also show, for comparison, the HQ phase transition 
by the Maxwell construction by the thin solid line.

\begin{wrapfigure}{r}{0.42\textwidth}
\includegraphics[width=.41\textwidth]{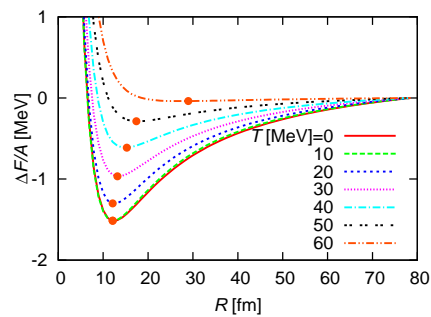}
\caption{
Droplet radius $R$ dependent free energy per baryon at $\rho_B=2 \rho_0$ for different temperatures. 
The quark volume fraction $(R/R_W)^3$ is fixed to be the optimal value at $T=0$ for each curve. 
The free energy is normalized by the value at $R\rightarrow\infty$.
Dots on each curve shows the energy minimum. 
}
\label{fig:stable}
\end{wrapfigure}
Compared with the zero temperature case, the mixed phase is restricted and 
the EOS gets close to that by the Maxwell construction at finite $T$,
though we properly apply the Gibbs conditions. 
The restriction of the mixed phase has been already demonstrated at $T=0$ due to the charge screening effect \cite{maru07}. 
We can see that the geometrical structure also becomes unstable due to the thermal effect. 
Figure~\ref{fig:stable} shows the free energies per baryon 
of the droplet structure at several values of temperature. 
The quark volume fraction $(R/R_W)^3$ is fixed to exclude the trivial $R-$ dependence. 
Here we use, for example, the optimal value of $(R/R_W)^3$ at $T=0$~MeV in every curve. 
We normalize them by subtracting the free energy at infinite 
radius,  
 $
  \Delta F = F(R)-F(R\rightarrow\infty),
 $
 to show the $R$ dependence clearly. 
The structure of the mixed phase is mechanically stable below $T \sim 60$ MeV, 
but the optimal value of the radius $R$ is shifted to the larger value as $T$ increases.
This behavior is opposite to the LG mixed phase of nuclear matter.
The reason is as follows: 
First, in the LG mixed phase, higher temperature causes
a reduction of surface tension, while we use a constant value for HQ mixed phase.
Second, higher temperature causes mixture of $\Sigma^-$ and $\Lambda$ in the hadron phase
of HQ mixed phase and reduces the Coulomb energy, while the LG mixed phase of
nuclear matter has small dependence of particle fraction on the temperature.
Thus the optimal value of $R$ is shifted to a larger value at finite temperature. 
Note that the extreme case such that $R$ (or $R_W$) $\rightarrow \infty$ corresponds 
to the Maxwell construction for bulk matter. 
In our formulation, the pasta structures disappear at $T \gtrsim 60$ MeV.

At zero temperature we have noticed the suppression of hyperon mixture
in the hadron phase \cite{maru07,maru08}.
At finite temperature, on the other hand, hyperons are mixed 
in the pure hadronic matter at lower density and in the mixed phase
at medium density.
This is analogous to the finite baryon density in the liquid phase
at low density nuclear matter discussed in Sec.~2.
At finite temperature, however, the mixture of hyperons in the mixed phase
is still suppressed \cite{yasutake} 
exactly by the same mechanism at zero temperature.

\section{Summary and concluding remarks}
We have investigated properties of mixed phase at the first-order phase transitions in nuclear matter
such as the LG mixed phase at low densities and the HQ mixed phase at high densities.
We have seen that pasta structures appear at finite temperatures as well as
at zero temperature.
For the LG mixed phase at low densities, the proton fraction $Y_p$ is a
crucial quantity.
If $Y_p\approx0.5$ the local proton fraction is almost constant and
baryon partial system behaves like a system with a single component.
Therefore the Maxwell construction is approximately applicable. 
However the finite-size effects induce the mechanical instability of pasta structures,
and reduce the region of the mixed phase.
At finite temperatures size of the structure becomes smaller.
This is due to the reduction of the surface tension between liquid and gas phases.
We have also studied the HQ phase transition at finite temperature. 
At finite temperature, the EOS gets close to that given by the Maxwell construction. 
It is due to the mechanical instability of the geometrical structure induced by the thermal effect.  
The temperature-dependence of the structure size has been opposite for the HQ and the LG mixed phases.
This is due to the different treatments on the surface tension and 
the different charged particles in the mixed phases.

In conclusion we emphasize that the existence of
pasta structures together with the 
finite-size effects are
common and general for the mixed phases at 
the first-order phase transitions in nuclear matter.
Particularly matter at finite temperature
exhibits various features which are
interesting and important for 
not only nuclear astrophysics but also thermodynamics.


%

\end{document}